# VLSI Implementation of Novel Class of High Speed Pipelined
# Digital Signal Processing Filter for Wireless Receivers

[1,2]Rozita Teymourzadeh, Yazan Samir Algnabi, Masuri Othman , Md Shabiul Islam
[2]Mok Vee Hong
[1]Institute of MicroEngineering and Nanoelectronics IMEN, VLSI Design Department,
Universiti Kebangsaan Malaysia, 43600 Bangi, Selangor, Malaysia
[2]Faculty of Engineering, Architecture and Built Environment,
Electrical & Electronic Engineering department, UCSI University, Kuala Lumpur, Malaysia

**Abstract:** The need for high performance transceiver with high Signal to Noise Ratio (SNR) has driven the communication system to utilize latest technique identified as over sampling systems. It was the most economical modulator and decimation in communication system. It has been proven to increase the SNR and is used in many high performance systems such as in the Analog to Digital Converter (ADC) for wireless transceiver. This research work presented the design of the novel class of decimation and its VLSI implementation which was the sub-component in the over sampling technique. The design and realization of main unit of decimation stage that was the Cascaded Integrator Comb (CIC) filter, the associated half band filters and the droop correction are also designed. The Verilog HDL code in Xilinx ISE environment has been derived to describe the proposed advanced CIC filter properties. Consequently, Virtex-II FPGA board was used to implement and test the design on the real hardware. The ASIC design implementation was performed accordingly and resulted power and area measurement on chip core layout. The proposed design focused on the trade-off between the high speed and the low power consumption as well as the silicon area and high resolution for the chip implementation which satisfies wireless communication systems. The synthesis report illustrates the maximum clock frequency of 332 MHz with the active core area of $0.308 \times 0.308$ $mm^2$. It can be concluded that VLSI implementation of proposed filter architecture is an enabler in solving problems that affect communication capability in DSP application.

## INTRODUCTION

The most popular ADC converters are realized based on the use of over sampling and sigma-delta ($\sum\Delta$) modulation techniques followed by decimation process (Ritoniemi *et al.*, 1994). Oversampled Sigma delta ($\sum\Delta$) modulator provides high resolution sample output in contrast to the standard Nyquist sampling technique. However at the output, the sampling process is needed in order to bring down the high sampling frequency and obtain high resolution. The CIC filter is a preferred technique for this purpose. A 3rd order sigma-delta ADC with a five bit data stream output is considered. The dynamic range required is 98 dB which can be achieved by efficient decimation system after applying oversampling technique.

The objective of this study is the on-chip implementation of novel class of high speed CIC filter designed for 5 bits data stream input, from 3rd order sigma delta modulator adapted for multi-standard wireless receiver. This filter performs both filtering of the out of band quantization noise and prevents excess aliasing introduced during sampling rate decreasing. Furthermore, in order to save power consumption and high speed implementation, computation complexity should be relaxed based on filter order and number of output bit.

For the first time in 1981, Hogenauer (1998) invented a new class of economical digital filter for decimation called a Cascaded Integrator Comb filter (CIC) or recursive comb filter. This filter worked with sampling frequency of 5 MHz. The introduced CIC filter did not required storage for filter coefficients and multipliers as all coefficients are unity (Park, 1990). Furthermore its on-chip implementation is efficient because of its regular structure consisting of three basic building blocks, minimum external control and less complicated local timing is required and its change factors is reconfigurable with the addition of a scaling circuit and minimal changes to the filter timing. It is also used to perform filtering of the out of band quantization noise and prevent excess aliasing introduced during sampling rate decreasing. Hence implementing high performance CIC filter will be issue of this research work.

Garcia *et al*. (1998) designed Residue Number System (RNS) for pipelined Hogenauer CIC. Compared to the two's complement design, the RNS based Hogenaur filter enjoys an improved speed advantage by approximately 54%. Similar structure by Meyer-Baese *et al*. (2005) has been implemented to reduce the cost in the Hogenauer CIC filter which shows that the filter can operate up to maximum clock frequency of 164.1 MHz on Altera FPLD and 82.64 MHz on Synopsys cell-based IC design.

Abu-Al-Saud and Stuber (2003), CIC filter structure was slightly modified to enhance its sample rate conversion performance at the expense of requiring a few extra computations. However, it is still necessary to design new algorithm in terms of high performance requirement. Later, Mortazavi *et al*. (2005) proposed recursive CIC filter with 120 dB stop-band attenuation and less than 0.0003 pass-band ripples. The CIC filter functionality was well described in bit-serial and bit-parallel arithmetic in this research work.

Wang *et al*. (2009), in 2009 presented principles of CIC filter which was used in energy calculations. They proposed a new structure of third order CIC decimation filter for electric power metering chip application. In the mentioned design, saving hardware resources and cost reduction was the main concern.

Similar work through the years 2005-2010 also presented (Shahana *et al*., 2007; Dolecek and Mitra 2008; Sun and Chen, 2008; Mortazavi *et al*., 2009), to show the efforts that engineers put to increase the CIC filter efficiency for four Generation (4G) filter applications. However, there are inadequate resources regarding high performance decimation filters and enhancement of CIC filter performance is still a challenging task.

Consequently, this study focused on the chip implementation of novel algorithm of five order high speed CIC filters with finite-precision effect and adaptive filter time varying systems. In order to reduce filter calculations and decrease active core area for 4G wireless communication applications, high speed adder was designed and implemented based on carry look-ahead adder (CLA) equations. ASIC realization of proposed CIC filter leads us to optimize filter functionality such as power consumption and active core area.

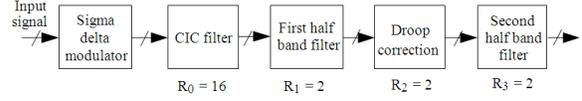

Fig. 1: Digital decimation process

Table 1: Decimation filters specifications

|  | Pass band (kHz) | Stop band (kHz) | Transition band (kHz) |
|---|---|---|---|
| CIC filter | 7.00 | 384.00 | 377.00 |
| First half band filter | 32.00 | 170.00 | 138.00 |
| Droop correction | 32.00 | 70.00 | 38.00 |
| Second half band filter | 21.77 | 26.53 | 4.76 |

**Development of a decimation filter:** The purpose of the CIC filter is twofold; firstly to remove filtering noise which could be aliased back to the base band signals and secondly to convert high sample rate m bit data stream at the output of the Sigma-delta modulator to n bit data stream with lower sample rate. This process is also known as decimation which is essentially performing the averaging and a rate reduction functions simultaneously. Figure 1 shows the decimation process using CIC filter.

The two half band filters (Brandt and Wooley, 1994) are used to reduce the remain sampling rate reduction to the Nyquist output rate. First half band filter and second half band filter make the frequency response flatter and sharper similar to the ideal filter frequency response. Droop correction filter is allocated to compensate pass band attenuation which is created by the CIC filter. A sample of the frequency response of overall decimation process is given as Table 1.

As shown in Table 1, the final frequency bandwidth of overall decimation system can satisfy audio application whereas 4G wireless communication requires wider band width and lower oversampling ratio with dynamic range of 98 dB. Hence it is the motivation to design novel class of high speed CIC filter to be applied in DSP block of transceiver.

**Principle of CIC filter structure:** The CIC filter consist of N stages of integrator and comb filter which are connected by a down sampler stage as shown in Fig. 1 in z domain. The CIC filter has the following transfer function:

$$H(z) = H_I^N(z).H_C^N(z) = \frac{(1-z^{-RM})^N}{(1-z^{-1})^N} = (\sum_{k=0}^{RM-1} z^{-k})^N \qquad (1)$$

Where:
$N$ = The number of stage
$M$ = The differential delay
$R$ = The decimation factor

In this project, *N*, *M* and *R* have been chosen to be 5, 1 and 16 respectively to avoid overflow in each stage.

N, M and R are parameters to determine the register length requirements necessary to assure no data loss. Equation 1 can be expressed as follows:

$$H(z) = \sum_{k=0}^{(RM-1)N} h(k)z^{-k} = \left[\sum_{k=0}^{RM-1} z^{-k}\right]^N \leq \left|\sum_{k=0}^{RM-1} z^{-k}\right|^N$$

$$\leq \left(\sum_{k=0}^{RM-1} |z|^{-k}\right)^N = \left(\sum_{k=0}^{RM-1} 1\right)^N = (RM)^N \quad (2)$$

From the Eq. 2, the maximum register growth/width, $G_{max}$ can be expressed as:

$$G_{max} = (RM)^N \quad (3)$$

In other words, $G_{max}$ is the maximum register growth and a function of the maximum output magnitude due to the worst possible input conditions (Hogenauer, 1998). If the input data word length is $B_{in}$, the Most Significant Bit (MSB) at the filter output, $B_{max}$ is given when M is equal to1:

$$B_{max} = [N \log_2 R + B_{in} - 1] \quad (4)$$

In order to reduce the data loss, normally the first stage of the CIC filter has maximum number of bit compared to the other stages. Since the integrator stage works at the highest oversampling rate with a large internal word length, decimation ratio and filter order increase which resulted in more power consumption and speed limitation.

**Proposed filter architecture:** The new algorithm of proposed CIC filter with dynamic range of 98 dB is investigated. The new algorithm was applied as linear adaptive filter. The filter is capable to involve both order-update and time-update recursions. The new adaptive filter distinguishes itself by virtue of order updates which are made possible by exploiting the time–shifting property of uniformly sampled temporal data. The designed filter is known collectively as fast algorithm because its computational complexity increases linearly with the number of adjustable parameter and more demanding in code terms.

**Truncation for low power and high speed:** Truncation means estimating and removing Least Significant Bit (LSB) to reduce the area requirements on chip and power consumption and also increase speed of calculation. Although this estimation and removing introduces additional error, the error can be made small enough to be acceptable for wireless communication systems. Figure 3 illustrates five stages of the CIC filter when is 25 bit $B_{max}$ so truncation is applied to reduce register width. Calculations and MATLB software helps to find word length in integrator and comb section respectively.

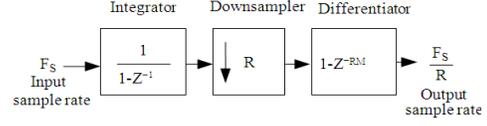

Fig. 2: One-stage of CIC filter block diagram

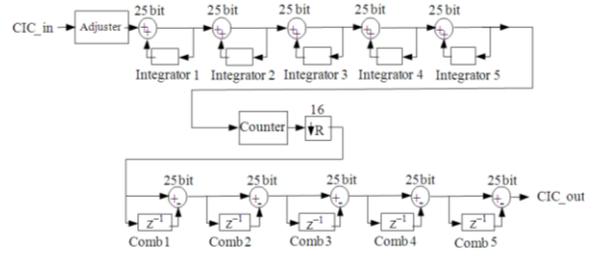

Fig. 3: Five-stages of truncated CIC filter

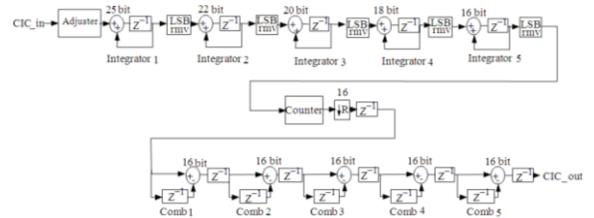

Fig. 4: Five-stage of truncated pipeline CIC filter

**New pipelining architecture:** The way to have high speed CIC filter is by implementing the pipeline filter architecture. Figure 4 shows pipeline CIC filter structure when truncation is also applied. This pipelining is different compared with normal pipeline structures. In the proposed CIC adaptive filter, no additional pipeline registers are used in integrator part. This is an important advantage of projected CIC filter because not only the system satisfies pipeline structure but also saves power consumption and reduces area on chip implementation. However, the hardware requirement is more relaxed compared with similar pipeline structures. The CIC decimation filter clock rate is determined by the first integrator stage that causes more propagation delay (Djadi *et al.*, 1994) than any other stage due to maximum number of bit.

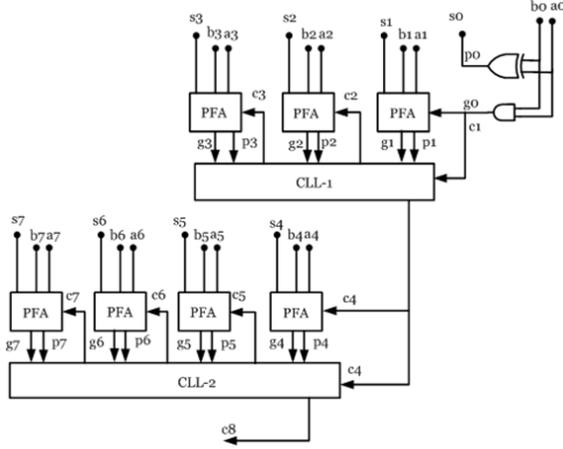

Fig. 5: The 8 bit proposed adder architecture

Hence it is possible to use a higher clock rate for a CIC decimation filter if a pipeline structure is used in the integrator stages, as compared to non-pipelined integrator stages. The clock rate in integrator section is R times higher than in the comb section.

Previously, the pipeline structure for CIC filter was applied just for integrator part since the maximum clock rate is determined by the integrator. The above architecture showed that the maximum throughput was increased by 60 MHz when the pipeline structure is used for all the CIC parts consisting of integrator, comb and down sampler.

**Proposed high speed adder:** The major technique to increase speed is by using proposed fast adder. This adder designed and implemented based on Carry Look-ahead Adder (CLA). The CLA is the fast adder which can be used for speeding up purpose but the disadvantage of the CLA adder is that the carry logic is gets quite complicated for more than 4 bits hence the proposed adder is introduced to replace as adder. This high speed is due to the carry calculation in CLA. In the ripple carry adder, the most significant bit addition has to wait for the carry to ripple through from the least significant bit addition. Therefore the carry of fast adder has become the focus of research (Ciletti, 2004).

The 8 bit designed adder architecture is shown in Fig. 5. Its block diagram consists of 2, 4 bit module which is connected. Each previous 4 bit calculates carry out for the next carry.

The new CIC filter in this study utilizes five fast adders in integrator parts. The maximum number of bit is 25 and it is decreased in the next stages. Hence it truncated respectively to 25, 22, 20, 18 and 16 bit in each adder, left to right. Notice that each 4 bit adder provides a group propagate and generate signal, which is used by the adder Logic block.

The group Propagate $P_G$ and Generate $G_G$ of a 4 bit adder will have the following expressions:

$$P_G = p_3 \cdot p_2 \cdot p_1 \cdot p_0 \qquad (5)$$

$$G_G = g_3 + p_3 \cdot g_2 + p_3 \cdot p_2 \cdot g_1 + p_3 \cdot p_2 \cdot p_1 \cdot g_0 \qquad (6)$$

The most important equations to obtain carry of each stage have been defined as below:

$$c_1 = g_0 + (p_0 \cdot c_0) \qquad (7)$$

$$c_2 = g_1 + (p_1 \cdot g_0) + (p_1 \cdot p_0 \cdot c_0) \qquad (8)$$

$$c_3 = g_2 + (p_2 \cdot g_1) + (p_2 \cdot p_1 \cdot g_0) + (p_2 \cdot p_1 \cdot p_0 c_0) \qquad (9)$$

$$
\begin{aligned}
c_4 = g_3 &+ (p_3 \cdot g_2) + (p_3 \cdot p_2 \cdot g_1) + \\
&(p_3 \cdot p_2 \cdot p_1 \cdot g_0) + (p_3 \cdot p_2 \cdot p_1 \cdot p_0 \cdot c_0)
\end{aligned} \qquad (10)
$$

Calculation of proposed adder is based on above equations. 8 bit high speed adder could be constructed continuing along in the same logic pattern, with the MSB carry-out resulting from OR and AND gates. The Verilog code has been written to implement addition. The projected adder Verilog code was downloaded to the Xilinx FPGA chip. From Xilinx ISE Virtex-5 synthesize report; it was found that the minimum clock period is 1.986 ns (Maximum Frequency is 503.449 MHz).

## RESULTS

After the sigma delta modulator, the sampling rate must be reduced to the Nyquist sampling rate. This is carried out in 4-stages. The first stage involves the reduction of the sampling frequency by the decimation factor of $R_0$. This is done by the proposed CIC filter. The remaining 3 stages involve the reduction of the sampling frequency by the decimation factor of $R_n$ which are carried out by the first half band, droop correction and the second half band respectively. Figure 6 illustrates the frequency response of the overall decimation filter when the decimation factor is 128.

Figure 7 creates the droop correction filter result. This filter designs a low-pass filter with pass-band having the shape of inverse of the CIC filter frequency response. Consequently it compensates the amplitude droop cause of the CIC filter.

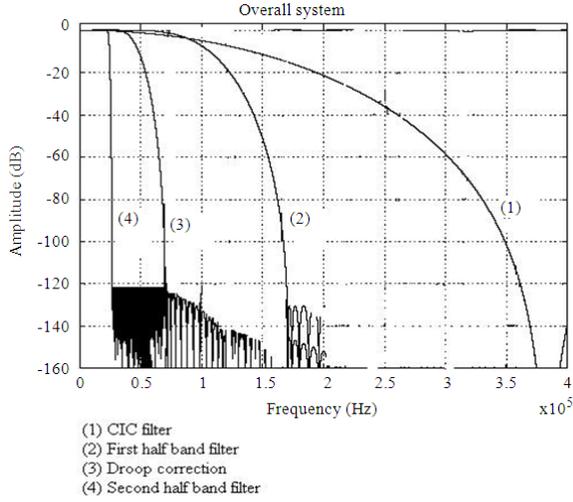

(1) CIC filter
(2) First half band filter
(3) Droop correction
(4) Second half band filter

Fig. 6: Frequency response of overall decimation filter

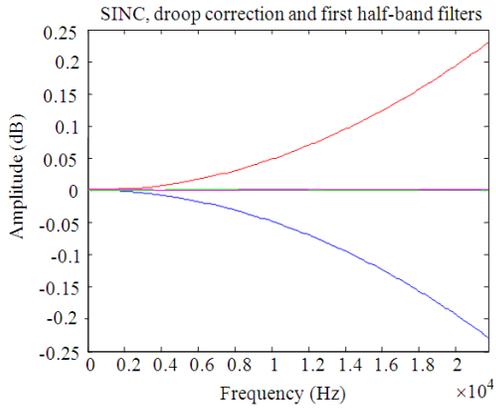

Fig. 7: Droop correction effect on frequency response

The designed novel class of CIC filter Verilog code was derived and simulated by MATLAB software. For a statistical analysis of its performance for the detection of sinusoidal signals in wideband noise, The SNR is illustrated 141.56 dB in sigma delta modulator output and it is increased to 145.35 dB in the decimation stages. To improve the SNR, word length of recursive CIC filter should be increased; whilst the speed of filter calculation is also decreased.

Table 2 shows the specification of proposed CIC filter and Fig. 8 creates the core layout of CIC filter on FPGA board.

As shown in Table 2, the synthesis report of Xilinx ISE environment indicates the maximum throughput of 330 MHz in filter specification. The dynamic range of proposed CIC filter is 98 dB which satisfies wireless communication transceiver systems (Grati *et al.*, 2002).

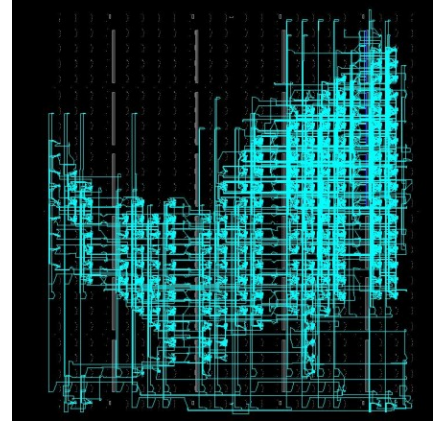

Fig. 8: Core layout on FPGA board

To optimize the projected CIC filter in terms of power consumption and active core area measurement ASIC implementation was performed successfully. Fig. 9 illustrates the active core layout of proposed CIC filter followed by Table 3 related to the filter specification in terms of power consumption and area measurment.

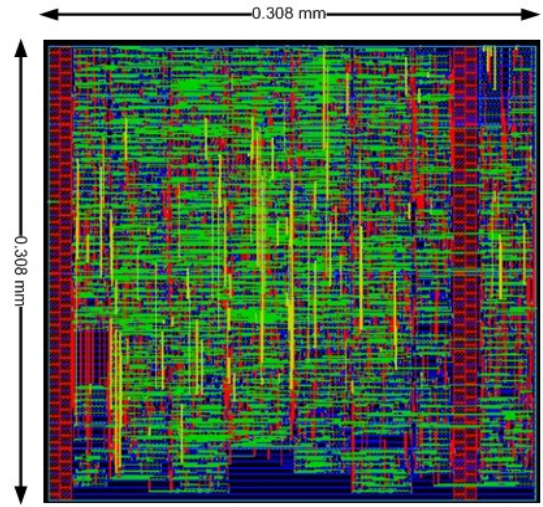

Fig. 9: ASIC active core layout of CIC filter

Table 2: High speed pipelined CIC filter characteristic

| HDL synthesis report | |
|---|---|
| **Design statistics** | |
| No. of IOs | 33.000 |
| No. of multiplexer | 2.000 |
| No. of Xor | 240.000 |
| No. of slices | 774.000 |
| **Timing report** | |
| Minimum period (ns) | 3.004 |
| Maximum frequency (MHz) | 332.930 |
| Minimum input arrival time before clock (ns) | 2.997 |
| Maximum output required time after clock (ns) | 2.775 |
| Total equivalent gate count for design | 5216.000 |
| Total memory usage (MB) | 66.000 |

Table 3: ASIC implementation result of CIC filter

| CIC filter specification | Silterra 0.18 µm | Mimos 0.35 µm |
|---|---|---|
| Active core area (mm$^2$) | 0.308×0.308 | 1.148×1.148 |
| Power consumption (mW) | 3.14 | 6.03 |

It was implemented on ASIC under the Silterra and Mimos technology library which compiled the proposed design core to perform gate-level synthesis and optimization. After compiling, the netlist is created. The estimated area and power consumption result of the high speed filter in Silterra 0.18 µm and Mimos 0.35 µm technology in maximum clock frequency is shown in Table 3. The implementation of new CIC filter was done successfully and the actual data was tested by design analyzer system on FPGA board in real time.

## DISCUSSION

The FPGA implementation of novel class of high speed low power Recursive CIC filters have been designed and investigated to fulfill the requirement of wireless transceiver applications. The novel CIC filter was obtained by its designed efficient architecture. To achieve high speed filter new technique was applied consequently. The special pipeline architecture, designing proposed fast arithmetic calculation unit and truncation lead us to have high speed novel CIC filter and upgrade the filter structure to achieve the maximum throughput of 332.93 MHz. The evaluation indicates that the pipelined proposed CIC filter is attractive due to high speed when both the decimation ratio and filter order are not high as stated in the Hogenauer Comb filter. Since the first stage of the CIC filter require maximum word length and also because of the recursive loop in its structure, the reduction in power consumption is limited by the throughput. Thus the truncation will reduce the power consumption and the number of calculation. The power consumption computed using CAD tools (Cadence and Synopsys) and 0.18 µm Silterra technology library. It was reported 3.14 mW power at maximum clock frequency. The ASIC core measurement calculated as 0.308×0.308 mm2 on 0.18 µm Silterra and 1.148×1.148 mm2 on 0.35 µm MIMOS technology library. The results prove high enlargement of CIC filter performance in trade-off between power and speed compare to the previous similar research work (Shahana *et al.*, 2007; Sun and Chen, 2008; Mortazavi *et al.*, 2009; Brandt and Wooley, 1994; Djadi *et al.*, 1994; Ciletti, 2003; Grati *et al.*, 2002).

## REFERENCES

Abu-Al-Saud, W.A. and G.L. Stuber,. Modified CIC filter for sample rate conversion in software radio software. *IEEE J. Sig. Process.*,vol 10(5): 152-154. 2003. DOI: : 10.1109/LSP.2003.810023

Brandt, P. B. and A.B. Wooley. A low-power, area-efficient digital filter for decimation and interpolation. *IEEE J. Solid-State Circ.*, vol. 29(6): 679-687.1994.DOI: 10.1109/4.293113.

Ciletti, D. M. *Advanced digital design with the verilog HDL*. Prentice Hall, Department of Electrical and Computer Engineering University of Colorado at Colorado Springs. 2004. ISBN: 812032756X. pp. 958-959.

Djadi, Y., T.A. Kwasniewski, C. Chan and V. Szwarc. A high throughput programmable decimation and interpolation filter. *Proceeding of International Conference on Signal Processing Applications and Technology*, vol. 2. pp: 1743-1748.1994. Carlton university, Canada.

Dolecek, G.J. & S.K. Mitra. On Design of CIC decimation filter with improved response. *Proceeding of the IEEE Conference on Communications, Control and Signal Processing*, pp:1072-1076.2008. StJulians. DOI: 10.1109/ISCCSP.2008.4537383

Garcia, A., U. Meyer-Baese and F. Taylor, Pipelined hogenauer CIC filters using field-programmable logic and residue number system. *Acoustics, Speech Signal Processing IEEE International Conference*, pp. 3085-3088.1998. DOI: 10.1109/ICASSP.1998.678178.

Grati, K., A. Ghazel and L. Lirida Naviner, Relaxed decimtion filter specifications for wireless transceivers. *Proceeding of the IEEE Conference Electronics Circuits and Systems*, pp. 565-569.2002. DOI: 10.1109/ICECS.2002.1046228

Hogenauer, E.B., An economical class of digital filters for decimation and interpolation. *IEEE Trans. Acoust. Speech Signal Process.*, l: 155-162. USA. 1998. ISSN: 0096-3518

Meyer-Baese, U., S. Rao, J. Ramirez and A. Garcia. Cost-effective hogenauer cascaded integrator comb decimator filter design for custom ICs. *IEE Elect. J.*, 41: 158-160. 2005. USA. DOI: 10.1049/el:20057000

Mortazavi, S.M., S.M. Fakhraie and O. Shoaei, A comparative study and design of decimation filter for high-precision audio data converters. *Proceeding of the 17 IEEE International Conference on MICROELE Chronics*, pp: 139-143. 2005. Iran. DOI: 10.1109/ICM.2005.1590055.

Mortazavi, S.M., S.R. Omam, S.M. Fakhraie and O. Shoaei. Experimental evaluation of different realizations of recursive CIC filters. *IEEE Conference on Electrical and Computer Engineering,* pp: 1056-1059. 2009. Canada. DOI: 10.1109/CCECE.2006.277385.

Park, S. *Principles of sigma-delta modulation for analog-to-digital converters*. Motorola Inc, APR8/D Rev., 1.1990.

Ritoniemi, T., E. Pajarre, S. Ingalsuo, T. Husu and V. Eerola *et al*. A Stereo audio sigma-delta AD-converter. *IEEE J. Solid-State Circ.*, 29: 1514-1523. 1994. DOI: 0018-9200/94.

Shahana, T.K., R.K. James, B.R. Jose, K. Poulose Jacob and S. Sasi, 2007. Polyphase implementation of non-recursive comb decimators for sigma-delta A/D converters. *IEEE International Conference on Acoustics, Speech and Signal Processing,* pp: 825-828. 2007. Tainan. DOI: 10.1109/EDSSC.2007.4450253.

Sun, H. and J. Chen. A down-sample design under special condition in high-speed all- digital system. *IEEE International Symposium on Intelligent Information Technology Application*, pp: 269-273. 2008. DOI:10.1109/IITA.2008.557.

Wang, Y., H.M. Yuan and W. Chen, 2009. Design of CIC filter and DFC used in energy metering IC. *Proceeding of the IEEE Conference on Industrial Electronics and Applications*, pp: 1270-1274.2009. China DOI: 10.1109/ICIEA.2009 5138406